\documentclass[twocolumn,showpacs,floatfix,superscriptaddress,amsmath,amssymb,pre]{revtex4}
\usepackage{mathrsfs}
\usepackage{txfonts}
\usepackage{amssymb}
\usepackage{graphicx}
\usepackage{hyperref}
\usepackage{ulem}
 \usepackage{overpic}
 \usepackage{psfrag}
\usepackage{tabularx}
\usepackage{array}
\usepackage{placeins}
\makeatletter

\begin{document}

\title{Entanglement entropy and massless phase in the antiferromagnetic
 three-state quantum chiral clock model}

\author{Yan-Wei Dai}
\affiliation{College of Materials Science and Engineering, Chongqing University,
Chongqing 400044, The People's Republic of China}
 \affiliation{Centre for Modern Physics,
Chongqing University, Chongqing 400044, The People's Republic of China}

\author{Sam Young Cho}
\altaffiliation{E-mail: sycho@cqu.edu.cn}
 \affiliation{Centre for Modern
Physics, Chongqing University, Chongqing 400044, The People's Republic of China}
 \affiliation{Department of Physics, Chongqing University, Chongqing 400044, The People's Republic of China}

\author{Murray T. Batchelor}
\altaffiliation{E-mail: batchelor@cqu.edu.cn}
 \affiliation{Centre for Modern
 Physics, Chongqing University, Chongqing 400044, The People's Republic of China}
\affiliation{Mathematical Sciences Institute and Department of
 Theoretical Physics, Research School of Physics and Engineering,
 Australian National University, Canberra ACT 2601, Australia}

\author{Huan-Qiang Zhou }
 \affiliation{Centre for Modern
 Physics, Chongqing University, Chongqing 400044, The People's Republic of China}
 \affiliation{Department of Physics, Chongqing University, Chongqing 400044, The People's Republic of China}

\begin{abstract}
 The von Neumann entanglement entropy is used to estimate the critical point $h_c/J \simeq 0.143(3)$ of the
 mixed ferro-antiferromagnetic three-state quantum Potts model
 $H = \sum_i [ J ( X_i X_{i+1}^{\,2} + X_i^{\,2} X_{i+1} ) - h\, R_i ]$, where 
 $X_i$ and $R_i$ are standard three-state Potts spin operators and 
 $J>0$ is the antiferromagnetic coupling parameter.
 This critical point value gives improved estimates for two Kosterlitz-Thouless transition points in the
 antiferromagnetic ($\beta < 0$) region of the $\Delta$--$\beta$ phase diagram of the three-state quantum chiral clock model,
 where $\Delta$ and $\beta$ are, respectively, the chirality and coupling parameters in the clock model.
 These are the transition points $\beta_c \simeq - 0.143(3)$ at $\Delta = \frac12$ between
 incommensurate and commensurate phases and $\beta_c \simeq - 7.0(1)$ at $\Delta = 0$ between
 disordered and incommensurate phases.
 The von Neumann entropy is also used to calculate the central charge $c$ of the underlying conformal field theory
 in the massless phase $h \le h_c$.
 The estimate $c \simeq 1$ in this phase is consistent with the known exact value at the particular point $h/J = -1$
 corresponding to the purely antiferromagnetic three-state quantum Potts model.
 The algebraic decay of the Potts spin-spin correlation in the massless phase is used to estimate the
 continuously varying critical exponent $\eta$.
\end{abstract}

\maketitle

\section{Introduction}

The $N$-state asymmetric or chiral clock model was originally introduced to provide a
simple description of monolayers adsorbed on rectangular substrates \cite{ostlund, huse}.
The clock models can be considered as discrete versions of the continuous XY model.
The asymmetry or chirality in the model hamiltonian induces incommensurate floating phases
with respect to the periodicity of the underlying lattice,
with commensurate-incommensurate phase transitions of the Kosterlitz-Thouless type \cite{KT,KT2}
corresponding to the melting of the incommensurate phase.
The model we consider here is the one-dimensional quantum version of the
three-state chiral clock model.
The most general one-dimensional three-state quantum chiral clock model considered by
Howes, Kadanoff and den Nijs~\cite{HKD} is defined by the hamiltonian
\begin{equation}
H = -\sum_{j=1}^\infty \left[ \cos a(p_j - \Delta_p) + \beta \cos a (\theta_{j+1}-\theta_j - \Delta_\theta) \right],
\label{clock1}
\end{equation}
where $a=2\pi/3$ and the variables $p_j, \theta_j$ take the three eigenvalues $0,1,2$. They obey the commutation relations
\begin{equation}
{\mathrm e}^{i a p_j} {\mathrm e}^{i a \theta_k} = \omega^{\delta_{jk}} {\mathrm e}^{i a \theta_k} {\mathrm e}^{i a p_j},
\label{comm}
\end{equation}
with $\omega={\mathrm e}^{i a}$.
We set the chiral parameters to $\Delta_p = \Delta_\theta = \Delta$.
The three-state quantum chiral clock model is a candidate for exhibiting non-Abelian bound states beyond Majorana fermions, 
for which the chiral interactions $\Delta$ play a key role~\cite{Fendley}.

This model has a rich phase diagram in terms of the parameters $\Delta$ and $\beta$,
which was originally mapped out using strong coupling series expansion techniques~\cite{HKD}.
The coupling parameter $\beta$ plays the role of inverse temperature.
For chirality parameter $\Delta=0$ and $\beta>0$ the model reduces to the three-state quantum Potts chain 
with purely ferromagnetic interactions~\cite{potts,wu}.
Recent work on the ferromagnetic three-state quantum Potts chain has been motivated by the connection to
topological phases and edge modes in $Z_3$ parafermion spin chains \cite{par1,par2,par3,par4,AF},
and has also manifested the relation between degenerate groundstates and spontaneous symmetry breaking~\cite{Su, Dai}.

On the other hand, for $\Delta=0$ and $\beta < 0$ the model reduces to the much less studied quantum
Potts model with mixed ferro-antiferromagnetic interactions.
Importantly, in the antiferromagnetic region of the phase diagram of the chiral clock model 
in the $\Delta$--$\beta$ plane, 
the line $\Delta=0$, $\beta < 0$ is dual to the line $\Delta=1/2$, $1/\beta < 0$~\cite{HKD},
so that results obtained for the mixed ferro-antiferromagnetic three-state quantum Potts chain
apply directly to the phase diagram of the more general antiferromagnetic chiral clock model at $\Delta=1/2$.
Using strong coupling series expansion analysis, 
Kosterlitz-Thouless transitions were identified~\cite{HKD} at the critical points 
$(0, \beta_c)$ and $(1/2,1/\beta_c)$, where $\beta_c = -10 \pm 5$.
These are, respectively, points E and G in the phase diagram in Fig.~2 of Ref.~\cite{HKD}.
The quantum critical point E is between disordered and incommensurate phases and
point G is between incommensurate and commensurate phases.
Other estimates for $\beta_c$ were obtained from the
quantum formulation of the mixed ferro-antiferromagnetic three-state quantum Potts model~\cite{HM,mf}.
Based on analysis of small chain sizes, a massless phase was identified for some critical field value $\beta_c$,
with $\beta_c < -5$~\cite{HM}.
The approximation $\beta_c \simeq -5$ was estimated using a mean-field renormalization group method \cite{mf}.

In this paper we use the von Neumann entanglement entropy to investigate quantum criticality in
the mixed ferro-antiferromagnetic quantum Potts model of relevance to the antiferromagnetic region of the
$\Delta$--$\beta$ phase diagram of the three-state quantum chiral clock model.
For conformally invariant one-dimensional critical quantum spin chains
entanglement entropy has been demonstrated to be a useful tool for calculating the central charge $c$
of the underlying conformal field theory.
More generally, various measures of entanglement have been demonstrated to be
a useful means for the detection and classification of quantum phase transitions~\cite{vidaletal,reviews}.
For quantum spin chains, the groundstate entanglement entropy of
a subsystem formed by contiguous $\ell$ sites of an infinite system,
with respect to the complementary subsystem, has the leading
behavior $S = \frac{c}{3} \log_2 \ell$ if the system is critical, 
or $S = \frac{c}{3} \log_2 \xi$ if the system is near critical, with correlation length 
$\xi$~\cite{Calabrese,Korepin,CC09}.

We numerically calculate the groundstate energy and wavefunction of the
mixed ferro-antiferromagnetic three-state quantum Potts model using the
infinite Matrix Product State (iMPS) representation with the infinite time-evolving block decimation (iTEBD) algorithm~\cite{vidal2}
in order to determine the critical point $\beta_c$ and
the central charge in the massless phase from the von Neumann entanglement entropy.
We also estimate the critical exponent $\eta$ of the spin-spin correlation in the massless phase.

The paper is arranged as follows.
In Section II we outline the relationship between the chiral clock model and the
related mixed ferro-antiferromagnetic three-state quantum Potts model.
In Section III we use the iMPS approach to calculate the von Neumann bipartite entanglement
entropy and obtain the critical coupling $\beta_c$.
The central charge of the underlying conformal field theory in the massless phase
is determined from the von Neumann entropy and correlation length in Section IV.
Results for the spin-spin correlation function, and thus the critical exponent $\eta$ in the massless region,
are given in Section V.
Concluding remarks are given in Section VI.

\section{Model hamiltonian}

\subsection{The three-state chiral clock model and the three-state quantum Potts model}
\label{relation}

We begin by writing the chiral clock hamiltonian (\ref{clock1}) in a different form in terms of the operators
\begin{equation}
Z_j = {\mathrm e}^{i a p_j}, \quad X_j = {\mathrm e}^{i a \theta_j},
\end{equation}
for which the commutation relations (\ref{comm}) become
\begin{equation}
Z_j X_k  = \omega^{\delta_{jk}} X_k Z_j,
\end{equation}
where we recall $\omega={\mathrm e}^{i 2\pi/3}$. The hamiltonian (\ref{clock1}) is then
\begin{equation}
2H = -\sum_{j=1}^\infty \left[ {\mathrm e}^{-i \phi} Z_j + {\mathrm e}^{i \phi} Z_j^{\,-1} +
\beta \left( {\mathrm e}^{i \phi} X_j X_{j+1}^{\,-1} + {\mathrm e}^{-i \phi} X_j^{\,-1} X_{j+1}\right) \right],
\label{clock2}
\end{equation}
with $\phi = a \Delta$. In terms of the usual Potts spin-operators
\begin{equation}
  Z_j= \left(\begin{array}{ccc}
                    1 & 0 & 0 \\
                    0 & \omega & 0 \\
                    0 & 0 & \omega^2
            \end{array} \right),
\quad
  X_j= \left(\begin{array}{ccc}
                    0 & 1 & 0 \\
                    0 & 0 & 1 \\
                    1 & 0 & 0
            \end{array} \right),
\end{equation}
acting at site $j$ of the infinite chain,
the model reduces to the three-state quantum Potts hamiltonian
\begin{equation}
 2H = -\sum_{j=1}^\infty \left[ R_j + \beta \left( X_j X_{j+1}^{\,2} + X_j^{\,2} X_{j+1}\right) \right],
\label{clock3}
\end{equation}
when $ \Delta = 0$. Here the Potts spin-operator $R_j$ is given as
\begin{equation}
 R_j = Z_j + Z_j^{\,-1}=\left( \begin{array}{ccc}
                    2 & 0 & 0 \\
                    0 & -1 & 0 \\
                    0 & 0 & -1
            \end{array} \right),
\end{equation}
with the identities $Z_j^3 = X_j^3 = 1$.

The hamiltonian we thus consider is defined by
\begin{equation}
 H = \sum_{j=-\infty}^{\infty}
  \left[ J \left( X_j X_{j+1}^{\,2} + X_j^{\,2} X_{j+1} \right) - h\, R_j \right],
 \label{ham1}
\end{equation}
where $J/h = - \beta >0$ is the antiferromagnetic interaction strength and
$h$ represents  the transverse field.
This mixed ferro-antiferromagnetic three-state quantum Potts model
has been studied by a variety of conventional techniques in both the
classical \cite{mixed1,mixed2,mixed3,mixed4,mixed5,mixed6} and
quantum formulations \cite{HKD,HM}.
This quantum formulation of the mixed ferro-antiferromagnetic Potts model
was studied by Herrmann and Martin \cite{HM}.
Based on analysis of small chain sizes, a massless phase was identified for some critical field value $h_c$, with $h_c < 0.2 J$.
The approximation $h_c \simeq 0.2 J$ has been estimated using a mean-field renormalization group method \cite{mf}.
These values are to be compared with the first estimate~\cite{HKD} $h_c/J = 0.1\pm^{0.10}_{0.03}$,
obtained from series analysis of the quantum version of the clock model.

\begin{figure}[h]
\begin{center}
 \begin{overpic}[width=0.4\textwidth]{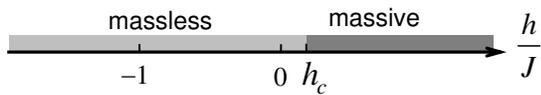}
\end{overpic}
\end{center}
\caption{
The phase diagram of hamiltonian (\ref{ham1}). 
The critical point $h_c$ separates massless and massive phases. 
For $h > 0$ the model is known as the mixed ferro-antiferromagnetic three-state quantum Potts model. 
The point $h/J=-1$ is the exactly solved antiferromagnetic three-state quantum Potts model. 
}
\label{fig1}
\end{figure}

Note that we consider only the case of antiferromagnetic coupling $J>0$ in the Potts hamiltonian (\ref{ham1}), 
corresponding to the antiferromagnetic region $\beta < 0$ in the phase diagram of the chiral clock model.
Given however, that the parameters appearing in the two Potts hamiltonians (\ref{clock3}) and (\ref{ham1}) are related 
by $\beta=-J/h$, we do not restrict ourselves to the values $h > 0$ of direct relevance to the 
antiferromagnetic region of the clock model.
Rather we also consider hamiltonian (\ref{ham1}) in the wider parameter space with $h < 0$. 
The phase diagram of this model is depicted in Fig.~\ref{fig1}.

\subsection{iMPS groundstate energy at the exactly solved point}
\label{energy}

 In order to study the mixed ferro-antiferromagnetic Potts model (\ref{ham1}) for an infinite-size chain, 
 the iTEBD method~\cite{vidal2} is used to obtain  
 the iMPS groundstate wavefunction $\left|\psi\right\rangle$ 
 and the groundstate energy for given parameter values.
 Fortunately, the model hamiltonian (\ref{ham1}) has been exactly solved at the particular field value $h/J=-1$,
 at which point the hamiltonian 
 is simply minus the hamiltonian of the  ferromagnetic three-state quantum Potts model at the self-dual critical point.
 For $h/J=-1$, the hamiltonian can be written simply in terms of the
 underlying Temperley-Lieb algebra \cite{berkcan} and
 exact results for the eigenspectrum can either be obtained by mapping to the equivalent
 spin-$\frac12$ XXZ chain or by solving the three-state model directly.
Using the latter approach, the groundstate energy per site for the infinite chain is~\cite{ADM}
\begin{equation}
 e_\infty = \frac43 - \frac{3\sqrt{3}}{2} - \frac{\sqrt{3}}{\pi} = -1.81607177\ldots .
\label{exact}
\end{equation}

For comparison with this result at $h/J=-1$, the iMPS groundstate energy per site has been calculated.
Specifically, we used first-order Trotter decomposition in the iTEBD algorithm, with an 
initial time step $dt = 0.1$ decreasing according to a power law until $dt = 10^{-6}$ 
as the groundstate is approached.
The numerical iMPS values are listed for several truncation dimensions $\chi$ in TABLE \ref{table}.
Note that the computational iMPS approach reproduces the exact result 
 (\ref{exact}) to $5$ significant figures already with truncation dimension $\chi=30$.
The significant figures for $\chi=150$ reach to $7$ digits.
This shows that the iMPS approach gives a reliable numerical result for the 
groundstate energy per site. 
We adapt this same approach for values of $h$ away from the exactly solved point.

\begin{table}[h]
\caption{\label{table}
iMPS estimates for the groundstate energy per site
of the antiferromagnetic three-state quantum Potts chain (\ref{ham1}) 
at the exactly solved point $h/J = -1$ with increasing truncation dimension $\chi$. 
Comparison is with the exact result (\ref{exact}).}
\vskip 2mm
\begin{ruledtabular}
\begin{tabular}{ccccccc}
& $\chi$ & 30 & 60 & 100  & 150 & \\
\hline
& $e_\chi$ &-1.81606688 & -1.81607095 & -1.81607153   & -1.81607168 & \\
\hline
& $\mathrm{error}$& $2.7 \times 10^{-6}$ & $4.5 \times 10^{-7}$   & $1.3 \times 10^{-7}$ & $5.0 \times 10^{-8}$ &\\
\end{tabular}
\end{ruledtabular}
\end{table}

\section{Entanglement entropy and quantum phase transition}
\label{sec_phase}

\begin{figure}[t]
\begin{center}
 \begin{overpic}[width=0.44\textwidth]{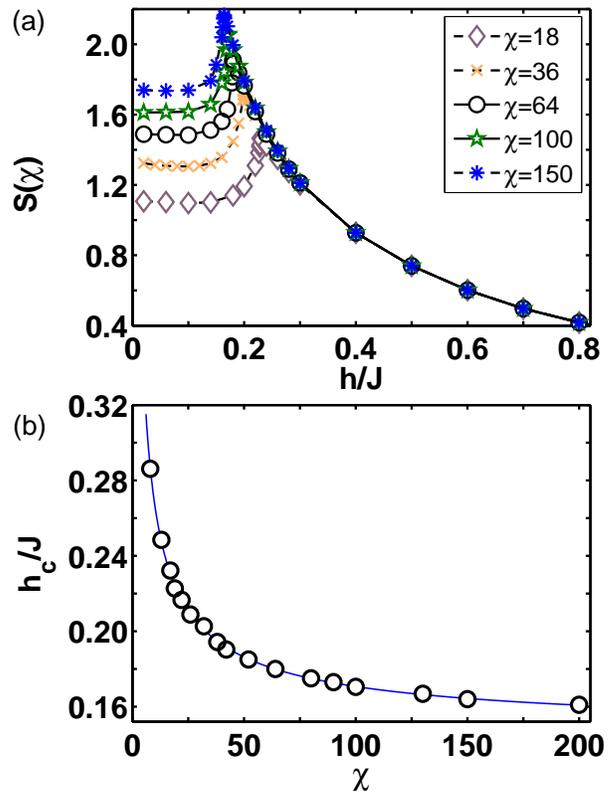}
\end{overpic}
\end{center}
\caption{(color online)
 (a) The von Neumann entanglement entropy $S(\chi)$ for the mixed ferro-antiferromagnetic three-state
 quantum Potts hamiltonian (\ref{ham1})
 as a function of the transverse field $h/J$ for increasing truncation dimension $\chi$.
 (b) The location of the peaks of the von Neumann entropies in (a)
 as a function of truncation dimension $\chi$.
 The solid line is the fitting function used
 to estimate the critical point $h_c/J = 0.143(3)$ in the thermodynamic ($\chi \to \infty$) limit (see text).
}
\label{fig2}
\end{figure}

 A quantum critical point in a given system can be detected by exploring thermodynamic properties for 
 which the system exhibits characteristic common singular behavior.
 It has been demonstrated recently that various entanglement measures are a useful means
 for detecting and classifying quantum phase transitions~\cite{vidaletal,reviews}.
 Especially, the von Neumann entropy has been shown to quantify quantum entanglement of a system
 and thus to detect singular behavior of quantum entanglement indicating the 
 occurrence of a quantum phase transition.

 In the iMPS approach the Schmidt decomposition coefficients of the bipartition between
 the semi-infinite chains $L(-\infty, \ldots, i)$ and $R(i + 1,\ldots,\infty)$,
 the elements of the diagonal matrix $\lambda^{[i]}_{\alpha_i}$ at site $i$
 can be used to evaluate the von Neumann entropy $S$~\cite{tagliacozzo,mukerjee}.
 In terms of the density matrix $\varrho = |\psi\rangle\langle \psi|$
 for the iMPS groundstate wavefunction $\left|\psi\right\rangle$,
 the von Neumann entanglment entropy
 is defined by $S = - \mathrm{Tr}[\varrho_L \log \varrho_L] = -
 \mathrm{Tr}[\varrho_R \log \varrho_R]$ where $\varrho_L$ and $\varrho_R$ are
 the reduced density matrices of the semi-infinite chains $L$ and $R$.
 In the iMPS representation, the von Neumann entropy is then calculated from~\cite{tagliacozzo,mukerjee}
 \begin{equation}
 S(\chi) =-\sum_{\alpha=1}^{\chi} \lambda^{2}_\alpha \log_{2}
  \lambda_{\alpha}^{2},
  \label{vonneumann2}
 \end{equation}
 where $\chi$ is the truncation dimension.

 The von Neumann entropy is plotted in Fig.~\ref{fig2}(a)
 as a function of the transverse field for increasing truncation dimension $\chi$.
 The von Neumann entropies are seen to exhibit a predominant peak structure
 with a singular point.
 The singular points $h_c(\chi)$ are indicative of a quantum phase transition.
 As the truncation dimension $\chi$ increases, the phase transition point $h_c(\chi)$ decreases.
 For an increment of $\chi$, the amplitude of the von Neumann entropy increases  for $h < h_c(\chi)$, with 
 little change for $h > h_c(\chi)$.
 This characteristic behavior of the von Neumann entropy implies that there are two distinct phases
 distinguished by the singular peak for a given truncation dimension.

 In order to estimate the quantum critical point $h_c$ in the thermodynamic $(\chi \rightarrow \infty)$ limit,
 the phase transition points are plotted
 as a function of the truncation dimension $\chi$ in Fig.~\ref{fig2}(b).
 The fitting function $h_c(\chi)/J = h_c/J + a \chi^b$
 is employed to perform the extrapolation of phase transition points~\cite{tagliacozzo}.
 With the numerical fitting coefficients $a=0.56(3)$ and $b=-0.66(3)$,
 we have obtained the estimate $h_c(\infty)/J \simeq 0.143(3)$ for the quantum critical point
 in the three-state quantum Potts hamiltonian (\ref{ham1}) with antiferromagnetic coupling $J$.
 According to the parameter relations between the three-state clock model (\ref{clock1})
 and the three-state Potts model in (\ref{ham1}),
 the corresponding critical point of the three-state quantum clock model 
 can be estimated as $\beta_c = -J/h_c(\infty) \simeq -7.0(1)$ for $\Delta=0$. 
 The duality symmetry for the three-state quantum clock model~\cite{HKD}, 
 i.e., the duality transformation $\beta \leftrightarrow 1/\beta$
 and $\Delta \leftrightarrow 1/2 -\Delta$, gives the corresponding critical point
 $\beta_c \simeq -0.143(3)$ for $\Delta = 1/2$.

\section {Massless phase and central charge}

\begin{figure}
 \includegraphics[width=0.44\textwidth]{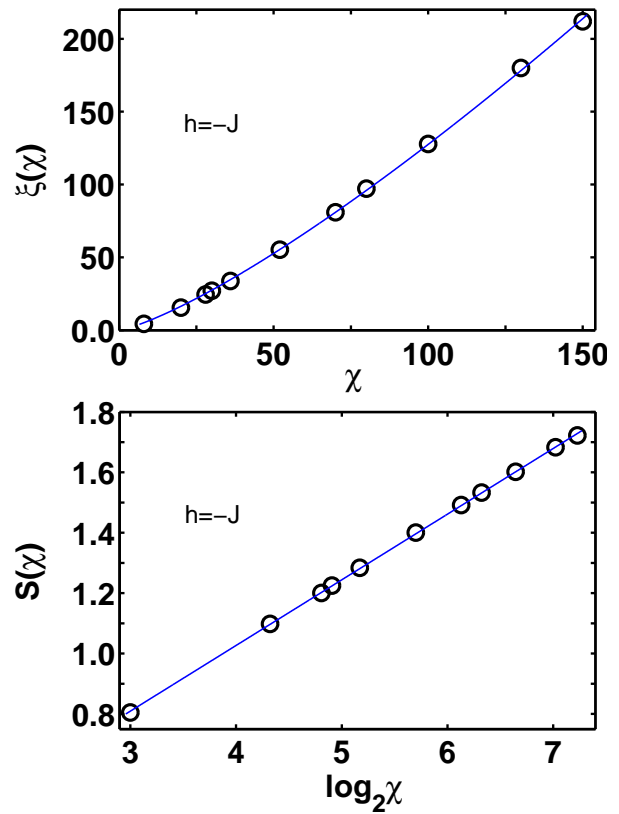}
\caption{(color online)
 Correlation length $\xi(\chi)$ and von Neumann entanglement entropy $S(\chi)$ as
 a function of the iMPS truncation dimension $\chi$ at the exactly solved point $h/J=-1$.
 The solid lines are the numerical fitting functions (see text).
}
  \label{fig3}
\end{figure}

\subsection{Criticality at the exactly solved point $h/J=-1$}
 \label{exact_point}

 As discussed in Sec.~\ref{energy}, the three-state Potts hamiltonian (\ref{ham1}) is an exactly solved model at $h/J = -1$.
 At this point the exact result $c=1$ for the central charge has been obtained from the
 finite-temperature thermodynamics derived from the Bethe Ansatz solution \cite{KM,albertini}.
 This is also the known value for the purely antiferromagnetic
 three-state quantum Potts model~\cite{baxter2,S,JS,I,note}.
 From the exact solution at this point it was established that the underlying conformal field theory
 is surprisingly given in terms of $Z_4$ parafermions.

 In the iMPS representation, for a critical groundstate, the central charge $c$
 can be studied and estimated from the scaling relations~\cite{tagliacozzo,mukerjee}
\begin{subequations}
\begin{eqnarray}
 S(\chi) &\sim& \frac{c \kappa}{6} \log_2 \chi,
 \label{S_chi}
 \\
 \xi(\chi) &\sim& a_\xi \, \chi^\kappa,
 \label{corr}
\end{eqnarray}
\end{subequations}
where $\kappa$ is a finite entanglement scaling exponent and $a_\xi$ is a constant.
The correlation length $\xi$ is defined
in terms of the largest and second largest eigenvalues of the transfer matrix
for a given truncation dimension in the iMPS representation by
$1/\xi(\chi)=\log_2 (\varepsilon_0(\chi)/\varepsilon_1(\chi))$.
 To obtain the central charge at $h/J=-1$ in our iMPS calculation,
 we plot the correlation length and the von Neumann entropy as a function of truncation dimension $\chi$
 in Fig.~\ref{fig3}.
 Both the correlation length and the von Neumann entropy increase as the truncation dimension $\chi$ increases.
 We first obtain the finite entanglement scaling
 exponent $\kappa$ by using a simple power law fitting on the correlation length in Eq.~(\ref{corr}).
 The fitting constants are $\kappa= 1.27(2)$ and $a_\xi=0.35(3)$.
 We then perform a best fit $S(\chi) = a + b \log_2 \chi$ for the von Neumann entropy to
 determine the coefficients $a=0.15(1)$ and $b=c\kappa/6 = 0.217(1)$.
 Combining these results gives the central charge estimate $c = 1.02(2)$.
 Our numerical result obtained from the von Neumann entropy 
 is thus consistent with the known exact result $c=1$ at the particular value $h/J=-1$.

\begin{figure}
\includegraphics[width=0.44\textwidth]{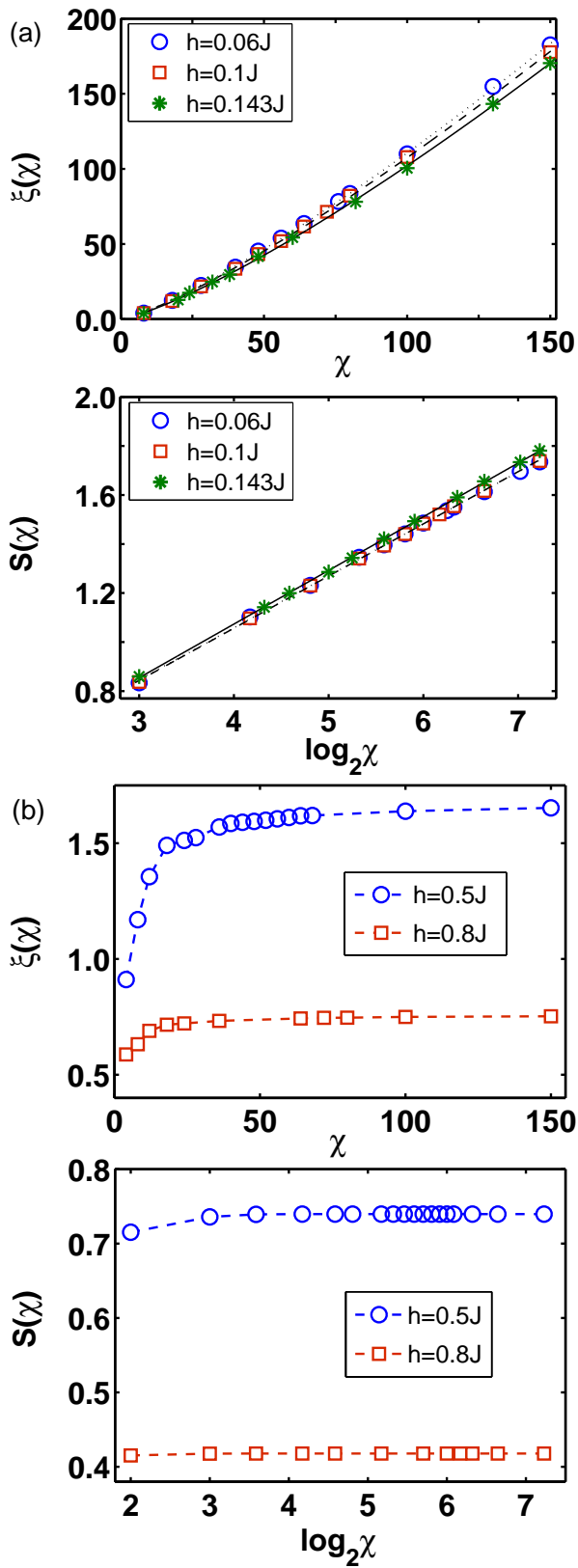}
\caption{(color online)
 Correlation length $\xi(\chi)$ and the von Neumann entanglement entropy $S(\chi)$ as
 a function of the iMPS truncation dimension $\chi$ in the (a) massless and (b) massive regions.
 The corresponding values of the field $h$ are indicated.
 Dashed lines for the plots in figure (a) correspond to the fitting functions discussed in the text.
}
  \label{fig4}
\end{figure}

 \subsection{Massless and massive phases}
\label{massless}

 As was noticed in Sec.~\ref{sec_phase},
 the amplitude of the von Neumann entropy increases with increasing truncation dimension $\chi$
 for $h \le h_c(\infty)$.
 Such behavior of the von Neumann entropy reveals a scaling behavior of the von Neumann entropy in a critical phase.
 More specifically, diverging behavior of the von Neumann entropy can characterize
 universality classes of a massless (critical) phase through the central charge $c$ of the underlying conformal field theory.
 In the iMPS representation, the central charge $c$ for $h \le h_c$
 can thus be studied and estimated from the scaling relations.
 Either massless phases or phases with a mass gap can then be readily distinguished by their scaling behavior in this approach.

 By using the scaling relations (\ref{S_chi}) and (\ref{corr}),
 we can thus obtain the central charge $c$ throughout the critical phase.
 Figure~\ref{fig4} shows plots of the correlation length $\xi(\chi)$ and the von Neumann entropy $S(\chi)$ 
 as a function of truncation dimension $\chi$ at various values of $h$.
 For $h \le h_c$ in Fig.~\ref{fig4}(a), the correlation length 
 and the von Nuemann entropy  diverge as the truncation
 dimension $\chi$ increases.
 In order to estimate first the finite entanglement scaling
 exponent $\kappa$, we have performed power law fitting on the correlation length in Eq.~(\ref{corr}).
 The numerical pairs of fitting constants are obtained as
 (i) $\kappa=0.124(2)$ and $a_\xi=0.35(7)$ for $h/J = 0.06$,
 (ii) $\kappa=1.24(1)$ and $a_\xi)=0.34(2)$ for $h/J = 0.1$,
  and (iii) $\kappa=1.26(2)$ and $a_\xi=0.30(3)$ for $h/J = 0.143$.
 These results show that for the chosen parameters, including the critical point $h_c(\infty) \simeq 0.143 J$,
 the finite entanglement scaling exponent $\kappa$ has a value very close to 
 the numerically obtained exponent at the exactly solved point $h/J=-1$ discussed in Sec. \ref{exact_point}.
 Next, to obtain the prefactor $b$ of the logarithmic divergence for the von Neumann entropy,
 we perform a best fit $S(\chi) = a + b \log_2 \chi$.
 The fitting constants obtained in this way are 
 (i) $a= 0.21(2)$ and $b=0.211(4)$ for $h/J = 0.06$,
 (ii) $a=0.20(1)$ and $b=0.212(2)$ for $h/J = 0.1$,
  and (iii) $a=0.19(1)$ and $b=0.218(1)$ for $h/J = 0.143$.
 Thus the relation $c = 6 b / \kappa$ gives
 the central charge estimates shown in TABLE \ref{tablec}.
 These results indicate that the central charge is almost certainly $c \simeq 1$ throughout the region $h \le h_c$
 and thus the system is in a critical regime, i.e., a massless phase for $h \le h_c$.
 In the language of the chiral Potts model, this is an incommensurate phase.

\begin{table}[h]
\caption{\label{tablec}
Estimates for the central charge $c$ in the massless phase of 
the mixed ferro-antiferromagnetic three-state quantum Potts model obtained from the von Neumann entanglement entropy 
at different values of $h$.}
\vskip 2mm
\begin{ruledtabular}
\begin{tabular}{ccccccc}
& $h/J$ & -1 & 0.06 & 0.1  & 0.143 & \\
& $c$& 1.02(2) & 1.02(3)   & 1.02(2) & 1.03(2) &\\
\end{tabular}
\end{ruledtabular}
\end{table}

 For $h > h_c$ in Fig.~\ref{fig4}(b), the correlation length 
 and the von Neumann entropy exhibit a simple saturation behavior as the truncation
 dimension increases.
 Such saturation behavior in both the correlation length and the von Neumann entropy
 indicate that the system is in a non-critical groundstate, i.e., a massive phase.
 In Fig.~\ref{fig4}(b), one can notice that the von Neumann entropy is bigger for $h = 0.5 J$ than for $h = 0.8J$.
 As can be seen in Fig.~\ref{fig2}(a), the von Neumann entropy becomes smaller as $h$ increases.
 Actually, for $h \rightarrow \infty$, the hamiltonian (\ref{ham1}) becomes $H \simeq -\sum_{-\infty}^{\infty} R_j $
 and then the groundstate is in a product state, which means that the von Neuman entropy becomes zero
 if the magnetic field $h \rightarrow \infty$.
 Consequently, our iMPS results show 
 distinct diverging or saturation behavior of the correlation and the von Neumann entropy 
 above and below the critical point $h_c$,
 characteristic of the massless phase for $h \le h_c$ or a massive phase for $h > h_c$, respectively.
Correspondingly the Potts hamiltonian (\ref{clock3}) has a massless phase for $\beta \leq \beta_c$ and a 
massive phase for $\beta > \beta_c$.

\section{Spin-spin correlations and critical exponent $\eta$}

 So far we have studied the quantum entanglement entropy and thus the characteristic behavior distinguishing
 the massless and massive phases in the three-state quantum Potts model (\ref{ham1})  
 and the corresponding three-state quantum clock model.
 In order to understand more about the physical nature of the massless and massive phases in these models,
 we investigate properties of the Potts spin-spin correlation defined by
\begin{equation}
 {\cal C}_{12}(|i-j|) = \left\langle X_i X_j^{\,2} \right\rangle.
\end{equation}
 In the iMPS approach, once the groundstate wavefunction is obtained,
 the expectation values of local or non-local physical operators can be calculated \cite{Su12}.
 In contrast to a finite-size lattice calculation,
 in principle, any lattice distance $r = |i-j|$ can then be considered for the Potts spin-spin correlation
 with the iMPS groundstate wavefunctions $|\psi\rangle$.

\begin{figure}
\includegraphics[width=0.44\textwidth]{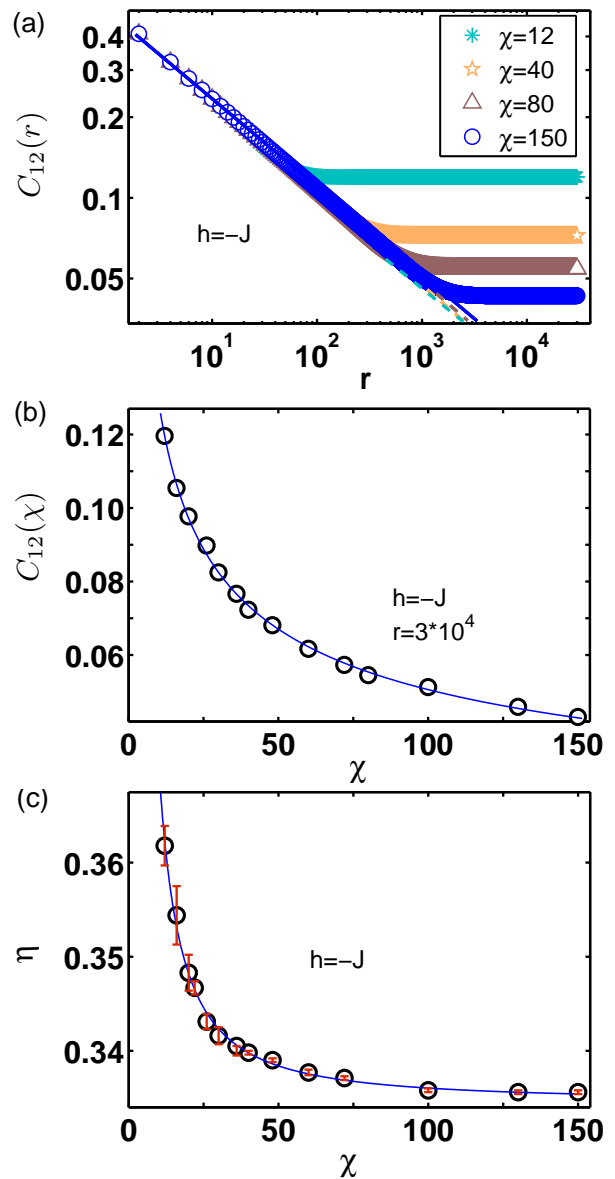}
\caption{(color online)
 (a) Potts spin-spin correlation
 $\mathcal{C}_{12}(r)$ as a function of the lattice distance $r=|i-j|$
 at field value $h/J=-1$.
 (b) Saturation value of Potts spin-spin correlation $\mathcal{C}_{12}(\chi)$ in (a)
  as a function of truncation dimension $\chi$ at $r = 3 \times 10^{4}$.
 (c) Spin correlation exponent $\eta$ as a function of truncation dimension.
  The exponent $\eta$ is given from the fitting function
  $C_{12}(r) = a_0 \, r^{-\eta}$ with the numerical constants $a_0$ and $\eta$
  for the algebraic decaying part in (a).
  The details are discussed in the text.
 } \label{fig5}
\end{figure}

 \subsection{Critical exponent $\eta$ at the exactly solved point $h/J=-1$}

 From the exact calculations,
 the exactly solved point $h/J=-1$ in the three-state quantum Potts model (\ref{ham1})
 is known to have the spin-spin correlation length critical exponent $\eta = 1/3$.
 In order to compare with this exact value at $h/J=-1$,
 in Fig.~\ref{fig5}(a),
 we plot the Potts spin-spin correlation  as a function of the lattice distance $r=|i-j|$
 with the iMPS groundstate wavefunction for various truncation dimensions.
 For all truncation dimensions,
 the spin-spin correlation shows an algebraic decay to its saturated value.
 The algebraic decaying part of the spin-spin correlation increases in distance 
 from a few hundreds to a few thousands of the lattice distance as the truncation dimension increases.
 Correspondingly, the saturation value decreases.
 From the observed trend,
 one may expect that the algebraic decaying range of the spin-spin correlation reaches an infinite lattice distance
 in the thermodynamic limit if the truncation dimension $\chi \to \infty$ 
 with the saturation value tending to zero.
 For a confirmation of this behavior, we consider a reasonably large lattice distance $r=|i-j| = 3 \times 10^{4}$
 and plot the saturation values of the Potts spin-spin correlation as a function of truncation dimension in Fig.~\ref{fig5}(b).
 The saturated value decreases as the truncation dimension $\chi$ increases.
 To quantify this behavior, we fit the function $C_{12}(3 \times 10^{4}) = a \chi^b +d$,
 which gives $a=0.32(3)$, $b=-0.40(7)$ and $d=1.56(1) \times 10^{-5}$.
 This indicates that the saturation value of the Potts spin-spin correlation tends to zero as $\chi \to \infty$.
 The observed saturation behavior is thus indeed a finite truncation effect.

 To estimate the critical exponent $\eta$ for the Potts spin-spin correlation in the thermodynamic limit,
 we consider the exponents of the algebraic decaying part of the spin-spin correlation.
 We performed a numerical fit to the algebraically decaying part with the function
 $C_{12}(r) = a_0 \, r^{-\eta}$ in Fig.~\ref{fig5}(a).
 The dashed lines in Fig.~\ref{fig5}(a) are fits with the parameter values
 (i) $a_0 = 0.533(2)$ and $\eta = 0.361(2)$ for $\chi = 12$,
 (ii) $a_0 = 0.5108(5)$ and $\eta = 0.3398(2)$ for $\chi = 40$,
 (iii) $a_0 = 0.5089(3)$ and $\eta = 0.3379(2)$ for $\chi = 80$,
 and (iv) $a_0 = 0.5070(4)$ and $\eta = 0.3356(2)$ for $\chi = 150$.
 These $\eta$ values show that the exponent of $\chi$ in the fitting function appears to be approaching
 the exact value $\eta_\infty = 1/3$ in the thermodynamic limit.
 To obtain $\eta_\infty$ in our iMPS calculation,
 we plot the estimates for $\eta$ for finite truncation dimensions in Fig.~\ref{fig5}(c).
 To extrapolate the critical exponent in the thermodynamic limit,
 we fit the function $\eta(\chi) = \eta_0 \chi^b + \eta_\infty$, with result $\eta_0=0.8(3)$, $b=-1.3(1)$ and 
 $\eta_\infty = 0.3346(9)$.
 This estimate at $h/J=-1$ is in excellent agreement with the exact value $\eta=1/3$.

\begin{figure}
\includegraphics[width=0.44\textwidth]{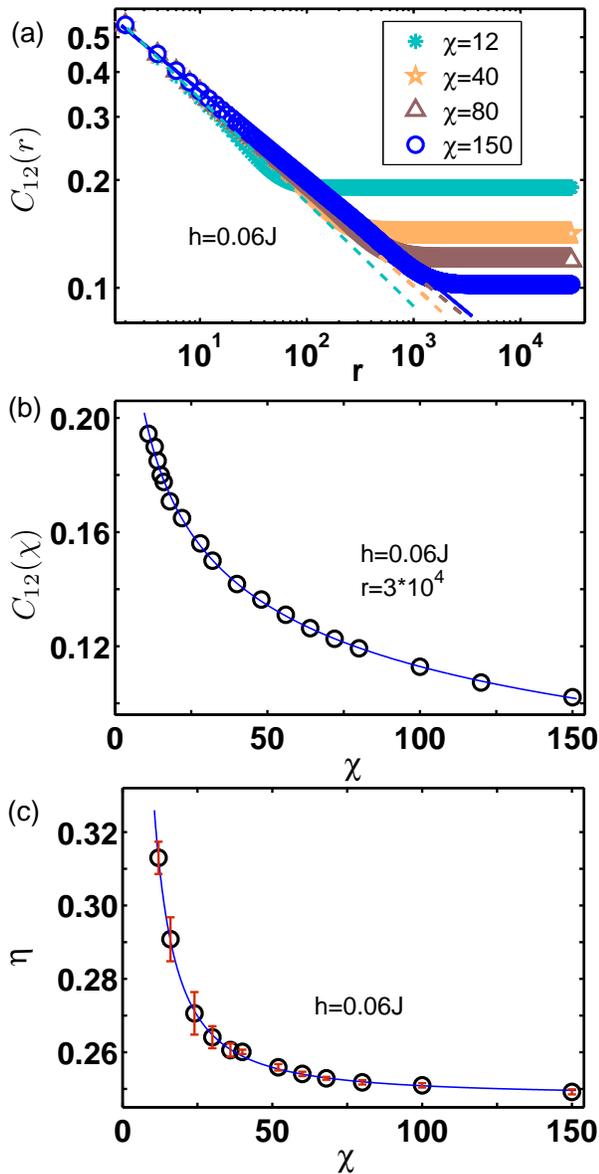}
\caption{(color online)
 (a) Potts spin-spin correlation
 $\mathcal{C}_{12}(r)$ as a function of the lattice distance $r=|i-j|$
 at field value $h/J=0.06$.
 (b) Saturation value of Potts spin-spin correlation $\mathcal{C}_{12}(\chi)$ in (a)
  as a function of truncation dimension $\chi$ at $r = 3 \times 10^{4}$.
 (c) Spin correlation exponent $\eta$ as a function of truncation dimension.
  The $\eta$ is given from the fitting function
  $C_{12}(r) = a_0 \, r^{-\eta}$ with the numerical constants $a_0$ and $\eta$
  for the algebraic decaying part in (a).
  The details are discussed in the text.
 } \label{fig6}
\end{figure}

 \subsection{Critical nature of the massless phase}

 As was shown in Sec.~\ref{massless}, the central charge $c\simeq1$ for $h \leq h_c$ indicates
 that the system defined by the hamiltonian (\ref{ham1}) is in a massless phase.
 In this subsection, we investigate the Potts spin-spin correlation in the massless phase.
 In order to compare with the detailed behavior of the spin-spin correlation
 at $h/J=-1$, in Fig.~\ref{fig6}(a),
 we plot the spin-spin correlation  as a function of the lattice distance $r=|i-j|$
 at $h/J=0.06$ for various truncation dimensions.
 For all truncation dimensions,
 the spin-spin correlations exhibit a similar behavior with those shown in Fig.~\ref{fig5}(a)
 for $h/J=-1$, i.e., they undergo a similar algebraic decay to their saturated values.
 Similar to the case $h/J=-1$, the saturation behavior is a finite truncation effect.
 To quantify this behavior, we plot
 the saturation values of the spin-spin correlation as a function of truncation dimension $\chi$
 for the lattice distance $r=|i-j| = 3 \times 10^{4}$ in Fig.~\ref{fig6}(b),
 which shows that the saturated value decreases as the truncation dimension $\chi$ increases.
 We fit the function $C_{12}(3 \times 10^{4}) = a \chi^b +d$,
 which gives $a=0.35(1)$, $b=-0.23(5)$ and $d=-0.01(3)$.
 Similarly to the case $h/J=-1$,
 this indicates that the saturation value of the spin-spin correlation tends to zero
 as $\chi \to \infty$.
 We then performed a numerical fit to the algebraically decaying part of the spin-spin correlation with the function
 $C_{12}(r) = a_0 \, r^{-\eta}$ in Fig.~\ref{fig6}(a).
 The dashed lines in Fig.~\ref{fig6}(a) are fits with the parameter values
 (i) $a_0 = 0.696(9)$ and $\eta = 0.313(4)$ for $\chi = 12$,
 (ii) $a_0 = 0.6381(4)$ and $\eta = 0.2601(6)$ for $\chi = 40$,
 (iii) $a_0 = 0.629(6)$ and $\eta = 0.2518(6)$ for $\chi = 80$,
 and (iv) $a_0 = 0.627(2)$ and $\eta = 0.2492(7)$ for $\chi = 150$.
 These estimates for $\eta$ indicate that the exponent of $\chi$ in the fitting function 
 decreases as the truncation dimension increases.
 To estimate $\eta_\infty$ 
 we plot the $\eta$ estimates for finite truncation dimensions in Fig.~\ref{fig6}(c).
 The extrapolation is performed with $\eta(\chi) = \eta_0 \chi^b + \eta_\infty$ where $\eta_0=2.6(5)$ and $b=-1.48(8)$, 
 which gives $\eta_\infty = 0.248(1)$.
 The critical exponent for $h/J=0.06$ in the thermodynamic limit 
 is thus $\eta_\infty =0.248(1)$.
 This exponent value clearly differs from the value at $h/J=-1$.

\begin{figure}
\includegraphics[width=0.44\textwidth]{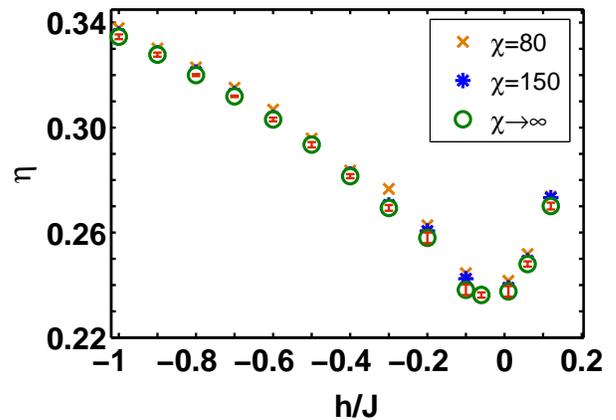}
\caption{(color online) Estimates for the Potts spin-spin correlation exponent $\eta$
 as a function of the transverse field $h/J$
 for the indicated values of the iMPS truncation dimension $\chi$.
 In the thermodynamic limit $\chi \rightarrow \infty$,
 the exponents are extrapolated in similar fashion to the estimates shown in Figs. \ref{fig5}(c)
 and \ref{fig6}(c).
 } \label{fig7}
\end{figure}

To further investigate the Potts spin-spin correlation in the massless phase 
we have performed similar calculations in the parameter range $ -J \leq h \leq 0.12J$
 and have observed similar finite truncation effects (the details are not presented here) 
 for the spin-spin correlations.
 To obtain the critical exponents $\eta$ in the thermodynamic $\chi \rightarrow \infty$ limit,
 a similar extrapolation has been performed.
 The estimates obtained in this way for the spin-spin correlation exponent $\eta$ 
 as a function of the field strength $h$ are 
 tabulated in TABLE \ref{tableeta} and plotted in Fig.~\ref{fig7}.
 This figure clearly shows that the critical exponent of the Potts spin-spin correlations
 is continuously varying
 and has a minimum value in the parameter range of the massless phase.
 The overall shape of this plot
 is similar with the estimates observed for the (Kosterlitz-Thouless) exponent $\eta$ for the
 purely ferromagnetic three-state quantum chiral chain~\cite{hamstudy}.
 
\begin{table}[h]
\caption{\label{tableeta}
Extrapolated estimates for the correlation length exponent $\eta$ in the massless phase of 
the mixed ferro-antiferromagnetic three-state quantum Potts model  
at different values of $h$.}
\vskip 2mm
\begin{ruledtabular}
\begin{tabular}{cccccc}
$h/J$ & $-1$ & $-0.9$ & $-0.8$ & $-0.7$ & $-0.6$ \\
$\eta_\infty$ & 0.3346(9) & 0.3278(8) & 0.3200(4) & 0.3119(3) & 0.3031(7) \\
\hline
~ & $-0.5$ & $-0.4$ & $-0.3$ & $-0.2$ & $-0.1$ \\
~ & 0.293(1) & 0.2815(8) & 0.269(1) & 0.258(2) & 0.238(2)\\
\hline
~ & $-0.06$ & 0.01 & 0.06 & 0.12 & ~ \\
~ & 0.236(1) & 0.237(2) & 0.248(1) & 0.270(1) & ~ \\
\end{tabular}
\end{ruledtabular}
\end{table}

 By combining exact results for the quantum sine-Gordon model
 with the Kosterlitz-Thouless theory of melting, such curves have been 
 predicted~\cite{schulz,ostlund,haldane} to have a minimum value of $2/q^2$ 
 and thus $2/9=0.222\ldots$ for $q=3$.
 However, Fig.~\ref{fig7} shows that our estimates may well be higher than
 the expected minimum value $\eta = 2/9$
 and also the expected (Kosterlitz-Thouless) exact value $\eta = 1/4$ at the critical point $h_c$.
 This discrepancy possibly originates from
 the difficulty of fitting finite-truncation dimension
 data to a power scaling law with a sufficient degree of accuracy.
 Then our data appears to show an overestimation of the Potts spin-spin correlation exponents
 in the thermodynamic limit due to the finite truncation effects in the iMPS approach.
 The value $c \simeq 1$ of the central charge is consistent with that of the Kosterlitz-Thouless type.

\begin{figure}[t]
\includegraphics[width=0.44\textwidth]{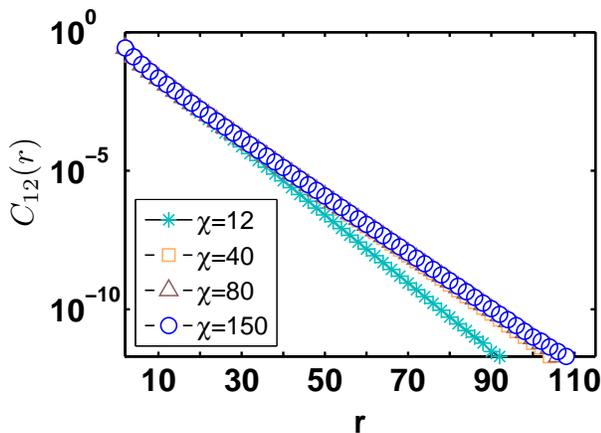}
\caption{(color online)
 Potts spin-spin correlation
 $\mathcal{C}_{12}(r)$ as a function of the lattice distance $r=|i-j|$
 in the massive phase at $h/J=0.5$.
 } \label{fig8}
\end{figure}

 In Fig.~\ref{fig8},
 to compare with the behavior of the Potts spin-spin correlations
 in the massless phase,
 we plot the spin-spin correlation
 as a function of the lattice distance $r=|i-j|$ for $h = 0.5J$
 in the massive phase.
 In contrast to the massless phase, the spin-spin correlation
 exponentially decays to zero in the massive phase for $h \geq h_c$.
 The log-linear plots show that the slope of the spin-spin correlation
 is readily saturated for the truncation dimension $\chi=150$.
 To quantify this behavior of the Potts spin-spin correlation
 for $h = 0.5J$, we fit the Potts spin-spin correlation for $\chi=150$ by
 using the function $C_{12}(r) = c_0 e^{- r/\xi_0}$
 with the fitting constant $c_0 = 0.18(1)$ and the Potts spin correlation length $\xi_0 = 4.21(1)$.
 As expected, when the magnetic field value approaches the critical point,
 the Potts spin correlation length $\xi_0$ becomes larger (the details are not presented here).
 Consequently, the spin-spin correlations in the three-state quantum Potts model (\ref{ham1}) 
 with antiferromagnetic coupling $J$ show
 characteristic behavior, i.e., algebraic decay to zero
 for the massless phase $h \leq h_c$ and exponential decay to zero
 for the massive phase $h > h_c$.

\section{Conclusion}

The von Neumann entanglement entropy has been demonstrated here to be an effective tool for estimating the
quantum critical point of the mixed ferro-antiferromagnetic three-state quantum Potts model (\ref{ham1}).
The critical point estimate $h_c/J \simeq 0.143(3)$ gives an improved estimate for two Kosterlitz-Thouless transitions
in the antiferromagnetic region of the $\Delta$--$\beta$
phase diagram of the quantum version of the three-state chiral clock model (\ref{clock1}).
The first transition point is located at $\beta_c = -J/h_c \simeq - 7.0(1)$ for $\Delta = 0$ between
disordered and incommensurate phases.
The second transition point is located at $\beta_c \simeq - 0.143(3)$ for $\Delta = 1/2$
between incommensurate and commensurate phases.
The latter point is deep within the phase diagram of the chiral clock model and follows
from the duality transformation $\beta \leftrightarrow 1/\beta$, $\Delta \leftrightarrow \frac12 - \Delta$~\cite{HKD}.

We also used the von Neumann entropy with the correlation length to calculate the central charge
of the underlying conformal field theory in the massless phase $h \le h_c$.
Our estimate $c \simeq 1$ indicates that the known exact value $c=1$ at the particular point $h/J = -1$
(the  antiferromagnetic three-state quantum Potts model) extends throughout the massless phase
of the mixed ferro-antiferromagnetic model, and thus into the massless phase of the three-state quantum chiral
clock model for $\beta < 0$.
This is an interesting feature of the $\Delta$--$\beta$ phase diagram of the three-state quantum chiral clock model. 
Previously it was demonstrated that the ferromagnetic three-state Potts model plays a key role in the ferromagnetic region $\beta > 0$ 
of the phase diagram~\cite{HKD}. 
It was not clear however, to what extent the antiferromagnetic three-state Potts model featured in the antiferromagnetic region $\beta <0$ 
of this phase diagram.
Here we have seen that although the antiferromagnetic three-state Potts model 
does not make a direct appearance in the $\Delta$--$\beta$ phase diagram -- 
in contrast to the ferromagnetic three-state Potts model -- it rather manifests itself indirectly through the 
value $c=1$ of the central charge in the massless incommensurate phase.
This is consistent with a recent DMRG study of the three-state quantum chiral clock model (in terms of different variables) where 
it has been shown that the value $c=1$ extends deep into the incommensurate phase~\cite{par3}.
We conclude by noting that the estimated continuously varying 
spin-spin correlation exponent $\eta$ in the parameter range of the massless phase
shown in Fig.~\ref{fig7} appears to be the typical shape
of Kosterlitz-Thouless exponent estimates in models
of this kind~\cite{hamstudy}.

{\it Acknowledgements.}
 MTB gratefully acknowledges support from Chongqing University and
 the 1000 Talents Program of China.
 This work is supported in part by the
 National Natural Science Foundation of China (Grant Numbers 11575037, 11374379 and 11174375) and
 Project Number 0903005203295 supported by the Fundamental Research Funds for the Central Universities.


\end{document}